\documentclass[12pt]{article}

\usepackage{scicite}

\usepackage{epsfig}
\usepackage{times}

\newenvironment{sciabstract}{%
\begin{quote} \bf}
{\end{quote}}

\newcounter{lastnote}
\newenvironment{scilastnote}{%
\setcounter{lastnote}{\value{enumiv}}%
\addtocounter{lastnote}{+1}%
\begin{list}%
{\arabic{lastnote}.}
{\setlength{\leftmargin}{.22in}}
{\setlength{\labelsep}{.5em}}}
{\end{list}}

\title{Fast variability of TeV $\gamma$-rays from the radio galaxy 
M\,87}

\author{
F.~Aharonian$^{1}$,
A.G.~Akhperjanian$^{2}$,
A.R.~Bazer-Bachi$^{3}$,
M.~Beilicke$^{4\ast}$,
W.~Benbow$^{1}$,\\
D.~Berge$^{1}$,
K.~Bernl\"ohr$^{1,5}$,
C.~Boisson$^{6}$,
O.~Bolz$^{1}$,
V.~Borrel$^{3}$
% ...
\\
\\
\normalsize{$^\ast$To whom correspondence should be addressed; E-mail: 
matthias.beilicke@desy.de}
}

\date{}

\begin{document} 


\maketitle 

{\footnotesize
\begin{enumerate}

\item Max-Planck-Institut f\"ur Kernphysik, P.O. Box 103980, D 69029
Heidelberg, Germany

\item Yerevan Physics Institute, 2 Alikhanian Brothers St., 375036
Yerevan, Armenia

\item Centre d'Etude Spatiale des Rayonnements, CNRS/UPS, 9 av. du Colonel
Roche, BP 4346, F-31029 Toulouse Cedex 4, France

\item Universit\"at Hamburg, Institut f\"ur Experimentalphysik, Luruper
Chaussee 149, D 22761 Hamburg, Germany

\item Institut f\"ur Physik, Humboldt-Universit\"at zu Berlin, Newtonstr.
15, D 12489 Berlin, Germany

\item LUTH, UMR 8102 du CNRS, Observatoire de Paris, Section de Meudon,
F-92195 Meudon Cedex, France

\item University of Durham, Department of Physics, South Road, Durham DH1
3LE, U.K.

\item Unit for Space Physics, North-West University, Potchefstroom 2520,
South Africa

\item Laboratoire Leprince-Ringuet, IN2P3/CNRS, Ecole Polytechnique,
F-91128 Palaiseau, France

\item Laboratoire d'Annecy-le-Vieux de Physique des Particules,
IN2P3/CNRS, 9 Chemin de Bellevue - BP 110 F-74941 Annecy-le-Vieux Cedex,
France

\item APC, 11 Place Marcelin Berthelot, F-75231 Paris Cedex 05,
France$^{\dag}$

\item Dublin Institute for Advanced Studies, 5 Merrion Square, Dublin 2,
Ireland

\item Landessternwarte, Universit\"at Heidelberg, K\"onigstuhl, D 69117
Heidelberg, Germany

\item Laboratoire de Physique Th\'eorique et Astroparticules, IN2P3/CNRS,
Universit\'e Montpellier II, CC 70, Place Eug\`ene Bataillon, F-34095
Montpellier Cedex 5, France

\item DAPNIA/DSM/CEA, CE Saclay, F-91191 Gif-sur-Yvette, Cedex, France

\item Laboratoire d'Astrophysique de Grenoble, INSU/CNRS, Universit\'e
Joseph Fourier, BP 53, F-38041 Grenoble Cedex 9, France

\item Institut f\"ur Astronomie und Astrophysik, Universit\"at T\"ubingen,
Sand 1, D 72076 T\"ubingen, Germany

\item Laboratoire de Physique Nucl\'eaire et de Hautes Energies,
IN2P3/CNRS, Universit\'es Paris VI \& VII, 4 Place Jussieu, F-75252 Paris
Cedex 5, France

\item Institute of Particle and Nuclear Physics, Charles University, V
Holesovickach 2, 180 00 Prague 8, Czech Republic

\item Institut f\"ur Theoretische Physik, Lehrstuhl IV: Weltraum und
Astrophysik, Ruhr-Universit\"at Bochum, D 44780 Bochum, Germany

\item University of Namibia, Private Bag 13301, Windhoek, Namibia

\item Universit\"at Erlangen-N\"urnberg, Physikalisches Institut,
Erwin-Rommel-Str. 1, D 91058 Erlangen, Germany

\item European Associated Laboratory for Gamma-Ray Astronomy, jointly
supported by CNRS and MPG

\end{enumerate}

\begin{itemize}

\item[$\dag$]{UMR 7164 (CNRS, Universit\'e Paris VII, CEA, Observatoire
de Paris)}

\item[\ddag]{now at Purdue University, Department of Physics, 525 
Northwestern Avenue, West Lafayette, IN 47907-2036, USA}

\end{itemize}

}


\begin{sciabstract}

\boldmath

The detection of fast variations of the TeV ($10^{12} \, \textrm{eV}$)
$\gamma$-ray flux, on time-scales of days, from the nearby radio galaxy
M\,87 is reported. These variations are $\sim 10$ times faster than that
observed in any other waveband and imply a very compact emission region
with a dimension similar to the Schwarzschild radius of the central black
hole. We thus can exclude several other sites and processes of the
$\gamma$-ray production. The observations confirm that TeV $\gamma$-rays
are emitted by extragalactic sources other than blazars, where jets are
not relativistically beamed towards the observer.

\unboldmath

\end{sciabstract}



So far the only extragalactic objects known to emit $\gamma$-radiation up
to energies of Tera electron volts ($1 \, \textrm{TeV} = 10^{12} \,
\textrm{eV}$) are blazars. These are active galactic nuclei (AGN) with a
plasma jet emanating from the vicinity of the black hole and pointing
close to the observer's line of sight. Due to the bulk relativistic motion
of the plasma in the jet the energy and luminosity of emitted photons are
boosted by relativistic effects, making blazars detectable up to TeV
energies.

The nearby radio galaxy M\,87 is located in the Virgo cluster of galaxies
at a distance of $\sim 16 \, \textrm{Mpc}$ ($z = 0.0043$) and hosts a
central black hole of $(3.2 \pm 0.9) \times 10^{9}$ solar masses
\cite{M87_BH_Mass}. The $2 \, \textrm{kpc}$ scale plasma jet
\cite{M87_Jet} originating from the centre of M\,87 is resolved at
different wavelengths (radio, optical and X-rays). The observed
inclination of the jet, at an angle of $\sim 30^{\circ}$ relative to the
observer's line of sight \cite{M87_JetAngle}, demonstrates that M\,87 is
not a blazar and hence would represent a new class of TeV $\gamma$-ray
emitters. M\,87 has also been suggested as an accelerator of the enigmatic
ultra-high-energy ($10^{20} \, \textrm{eV}$) cosmic rays
\cite{M87_UHECR_2,M87_SPB1}. Previously, weak evidence for $E > 730 \,
\textrm{GeV}$ $\gamma$-ray emission from M\,87 in 1998/1999 with a
statistical significance of $4.1$~standard deviations was reported by the
High Energy Gamma Ray Astronomy (HEGRA) collaboration \cite{HEGRA_M87_1}.  
No significant emission above $400 \, \textrm{GeV}$ was observed by the
Whipple collaboration \cite{M87_Whipple} from 2000-2003.

The observations reported here were performed with the High Energy
Stereoscopic System (H.E.S.S.) located in Namibia. H.E.S.S. is an array of
four imaging atmospheric-Cherenkov telescopes used for the measurement of
cosmic $\gamma$-rays of energies between $100 \, \textrm{GeV}$ and several
$10 \, \textrm{TeV}$, see \cite{HESS_Status} for more details. The
observations of M\,87 were performed between 2003 and 2006 yielding a
total of $89$ hours of data after quality selection cuts. After
calibration \cite{HESS_Calibration}, the H.E.S.S. standard analysis was
applied to the data using hard event selection cuts \cite{HESS_HardCuts}.
More information about the standard analysis, as well as a more recent,
alternative analysis technique \cite{ModelAnalysis} which gives consistent
results, can be found in \cite{SOM}.

An excess of $243$ $\gamma$-ray events is measured from the direction of
M\,87 in the whole data-set, corresponding to a statistical significance
of $13$~standard deviations, establishing M\,87 as a TeV $\gamma$-ray
source (Fig.~1). The position of the excess (Right Ascension $\alpha$ and
Declination $\delta$) was found to be $\alpha = 12^{\textrm{h}}
30^{\textrm{m}} 47.2^{\textrm{s}} \pm 1.4^{\textrm{s}}$, $\delta =
+12^{\circ} 23' 51'' \pm 19''$ (J2000.0). This is, within the quoted
statistical error and the systematic pointing uncertainty of the H.E.S.S.  
telescopes ($\sim 20''$ in both the Right Ascension and Declination
directions) compatible with the nominal (radio)  position
\cite{M87_Position} of the nucleus of M\,87 ($\alpha = 12^{\textrm{h}}
30^{\textrm{m}} 49.4^{\textrm{s}}$, $\delta = +12^{\circ} 23' 28''$).
Considering the angular resolution of H.E.S.S., the source is consistent
with a point-like object with an upper limit for a Gaussian
surface-brightness profile of $3$ arcmin ($99.9\%$ confidence level). At
the distance of M\,87 ($16 \, \textrm{Mpc}$) this corresponds to a radial
extension of $13.7 \, \textrm{kpc}$ which can be compared with the
large-scale structure of M\,87 as seen at radio wavelengths (Fig.~1). A
$\sim 10^{6}$ times stronger constraint on the size of the TeV emission
region is deduced from the observed short-term flux variability, as shown
below.

The differential energy spectra obtained for the 2004 and 2005 data sets
(Fig.~2) are both well fit by a power-law function $\textrm{d}N /
\textrm{d}E \propto E^{-\Gamma}$. The spectrum measured in 2005 is found
to be hard ($\Gamma \sim 2.2$) and reaches beyond $10 \, \textrm{TeV}$
with an average $\gamma$-ray flux of a factor of $\sim 5$ higher than in
2004.

The total $\gamma$-ray flux above $730 \, \textrm{GeV}$ (Fig.~3) for the
individual years from 2003 to 2006 indicatates variability on a yearly
basis \cite{M87_TEXAS} corresponding to a statistical significance of
$3.2$~standard deviations, being derived from a $\chi^{2}$ fit of a
constant function. The variability is confirmed by a Kolmogorov test
comparing the distribution of photon arrival times to the distribution of
background arrival times yielding a statistical significance for
burst-like (non-constant) behaviour of the source of $4.5$~standard
deviations. Surprisingly variability on time-scales of days (flux
doubling) was found in the high state data of 2005 (Fig.~3, upper panel)
with a statistical significance of more than $4$~standard deviations. This
is the fastest variability observed in any waveband from M\,87 and
strongly constrains the size of the emission region of the TeV
$\gamma$-radiation, which is further discussed below. No indications for
short-term variability were found in the data of 2003, 2004 and 2006,
which is not unexpected given the generally lower statistical
significances of the $\gamma$-ray excesses in those years.

These observational results (location, spectrum \& variability) challenge
most scenarios of very-high-energy $\gamma$-ray production in
extragalactic sources. Although the luminosity ($\approx 3 \times 10^{40}
\, \textrm{erg/s}$) of TeV $\gamma$-rays is quite modest and does not
cause any problems with the global energy budget of the active galaxy
M\,87, several models can be dismissed. The upper limit on the angular
size of $\sim 3$~arcmin ($13.7 \, \textrm{kpc} \approx 4.3 \cdot 10^{22}
\, \textrm{cm}$) centred on the M\,87 nucleus position already excludes
the core of the Virgo cluster \cite{M87_CRs} and outer radio regions of
M\,87 as TeV $\gamma$-ray emitting zones. Further, the observed
variability on time-scales of $\Delta t \sim 2$~days requires a very
compact emission region due to the light-crossing time. The characteristic
size is limited to $R \leq c \cdot \Delta t \cdot \delta \approx 5 \times
10^{15} \, \delta \, \textrm{cm} \approx 5 \times \delta \, R_{s}$, where
$\delta$ is the relativistic Doppler factor \cite{DefDopplerFactor} of the
source of TeV radiation and $R_{s} \approx 10^{15} \textrm{cm}$ is the
Schwarzschild radius of the M\,87 supermassive black hole (see below). For
any reasonable value of the Doppler factor (i.e. $1 < \delta < 50$, as
used in the modelling of TeV $\gamma$-ray blazars), this implies a drastic
constraint on the size of the TeV $\gamma$-ray source which immediately
excludes several potential sites and hypotheses of $\gamma$-ray
production. First of all this concerns the elliptical galaxy M\,87
\cite{M87_CRs} and the $\gamma$-ray production due to dark matter
annihilation \cite{M87_Neutralino}. The most obvious candidate for
efficient particle acceleration \cite{M87_JetsLimit}, namely the entire
extended kiloparsec jet, is also excluded. Although compatible with the
TeV source position, even the brightest knot in the jet (knot~A) appears
excluded with its typical size of the order of one arcsec (about $80 \,
\textrm{pc} \approx 2.5 \cdot 10^{20} \, \textrm{cm}$) resolved in the
X-ray range \cite{KnotA_Xrays}.


An interesting possibility would be the peculiar knot (HST-1) in the jet
of M\,87 (see supporting online text and Fig.~S2), a region of many
violent events, with X-ray flares exceeding the luminosity of the core
emission \cite{HST-1_Xrays} and super-luminal blobs being detected
downstream. Modelling the high-energy radiation properties of this region
(by synchrotron and inverse-Compton scenarios), several authors favour
sizes in the range of $0.1$ to $1 \, \textrm{pc}$ (for moderate values of
the Doppler factor ranging between 2 and 5)  \cite{M87_DopplerFactors,
M87_DopplerFactors2, HST-1_Xrays}. But, formally there is no robust lower
limit on the size of HST-1, therefore we cannot exclude HST-1 as a source
of TeV $\gamma$-rays. However, it would be hard to realize the short-term
variability of the TeV $\gamma$-ray emission in relation to HST-1, at
least within the framework of current models. While the size of the
$\gamma$-ray production region does not exceed $R \leq 5 \times 10^{15}
\delta \, \textrm{cm}$, the location of HST-1 along the jet at
$0.85$~arcsec from the nucleus, which corresponds to $d \approx 65 \,
\textrm{pc}$, implies that the energy would be channelled from the central
object into the $\gamma$-ray production region within an unrealistically
small opening angle $\sim R/d \approx 1.5 \times 10^{-3} \delta$ degree.

The only remaining and promising possibility is to conclude that the site
of TeV $\gamma$-ray production is the nucleus of M\,87 itself
\cite{M87_Nucleus}. In contrast to the established TeV $\gamma$-ray
blazars, the large scale jet of M\,87 is seen at a relatively large jet
angle ($\theta \sim 30^{\circ}$) which suggests a quite modest Doppler
boosting of its radiation. Nevertheless, due to the proximity of M\,87,
both leptonic \cite{M87_UC} and hadronic \cite{M87_SPB1, M87_SPB2} models
predicted detectable TeV $\gamma$-ray emission. However, these scenarios
typically produce a soft energy spectrum of TeV $\gamma$-rays, clearly in
contrast to the hard spectrum measured by H.E.S.S. Leptonic models can be
adapted in various ways to match the new results. Within
synchrotron-self-Compton (SSC)  scenarios \cite{SSC_Model}, one method is
to consider the possibility of differential Doppler-boosting in the jet
near the core region, a phenomenon clearly expected in the jet formation
zone which extends over less than $0.1 \, \textrm{pc}$ from the nucleus
\cite{M87JetFormationZone}. Emitting plasma blobs of small sizes with
Doppler factors between 5 and 30 and magnetic fields well below
equipartition can account for the observed TeV $\gamma$-ray emission. An
additional flux contribution from inverse-Compton scattering of background
photons, coming from scattered disk emission or from dust, can further
reduce the range of Doppler factors towards moderate values.

The TeV $\gamma$-ray photons (independent of their production mechanism)
might be absorbed by the pair absorption process $\gamma_{\textrm{TeV}} +
\gamma_{\textrm{IR}} \rightarrow e^{+} e^{-}$ on the local infrared (IR)
radiation field in the TeV $\gamma$-ray emission region. Since no
signature for an absorption can be identified in the energy spectrum up to
$10 \, \textrm{TeV}$, one can derive an upper limit on the luminosity of
the infrared radiation field at $0.1 \, \textrm{eV}$ (corresponding to a
wavelength of approximately $10$ micron, most relevant for absorption of
$10 \, \textrm{TeV}$ $\gamma$-rays) to be $L(0.1 \, \textrm{eV})  \leq 3.6
\times 10^{38} (R / 10^{15} \, \textrm{cm}) \, \textrm{erg} / \textrm{s}$,
where $R$ is the size of the TeV $\gamma$-ray emission region. Such a low
central IR radiation luminosity supports the hypothesis of an
advection-dominated accretion disk (i.e. an accretion disk with low
radiative efficiency) in M\,87 \cite{M87_ADAF} and generally excludes a
strong contribution of external inverse-Compton emission on IR light to
the TeV $\gamma$-ray flux.
 
If one accepts the hypothesis that protons can be accelerated as high as
$10^{20} \, \textrm{eV}$ in jets of radio galaxies, then (hadronic)  
proton synchrotron models \cite{M87_SPB1, M87_SPB2} can not be excluded
considering the presented data. An alternative $\gamma$-ray production
mechanism is curvature radiation of ultra-high-energy protons in the
immediate vicinity of the supermassive black hole. This novel mechanism
can simultaneously explain both the hard spectrum and fast variability of
the observed TeV $\gamma$-ray emission. Rapidly rotating black holes
embedded in externally supported magnetic fields can generate electric
fields and accelerate protons to energies up to $10^{20} \, \textrm{eV}$
\cite{BH_UHECR1, BH_UHECR2, BH_Gammas}. Assuming that acceleration of
protons takes place effectively within $3$ Schwarzschild radii $R_{s}$,
and if the horizon threading magnetic field is not much below $10^{4} \,
\textrm{G}$, one should expect $\gamma$-ray radiation due to proton
curvature radiation extending to at least $10 \, \textrm{TeV}$ (the
electron curvature radiation is less likely because of severe energy
losses even in a tiny component of an irregular magnetic field). No
correlation with fluxes at other wavelengths is expected in this model.
Although the size of the $\gamma$-ray production region, $R \sim 3 R_{s}
\sim 3 \times 10^{15} \, \textrm{cm}$ perfectly matches the observed
variability scale, and the model allows extension of the $\gamma$-ray
spectrum to $10 \, \textrm{TeV}$ without any significant correlation at
other wavelength, the main problem of the model is the suggested magnetic
field. It is orders of magnitude larger than the B-field expected from the
accretion process, given the very low accretion rate as it follows from
the bolometric luminosity of the core as well as the estimates of the
power of the jet in M\,87.

In summary, the time-scale of the short-term variability of the TeV
$\gamma$-rays is in the order of the light crossing time of the black hole
(located at the center of M\,87), which is a natural time-scale of the
object. Therefore, the results reported here give clear evidence for the
production of TeV $\gamma$-rays in the immediate vicinity of the black
hole of M\,87.


\bibliography{scibib}

\bibliographystyle{Science}


\begin{scilastnote}

\item Acknowledgement. The support of the Namibian authorities and of the
University of Namibia in facilitating the construction and operation of
H.E.S.S. is gratefully acknowledged, as is the support by the German
Ministry for Education and Research (BMBF), the Max Planck Society, the
French Ministry for Research, the CNRS-IN2P3 and the Astroparticle
Interdisciplinary Programme of the CNRS, the U.K. Particle Physics and
Astronomy Research Council (PPARC), the IPNP of the Charles University,
the South African Department of Science and Technology and National
Research Foundation, and by the University of Namibia. We thank D.~Harris
for providing the Chandra X-ray light curve of the M\,87 nucleus.

\end{scilastnote}

\subsection*{Supporting Online Material}
www.sciencemag.org \\
Materials and Methods \\
Supporting text \\
Figs.~S1, S2 \\
Tab.~S1 \\
References \\



\clearpage

\begin{figure}

\epsfig{file=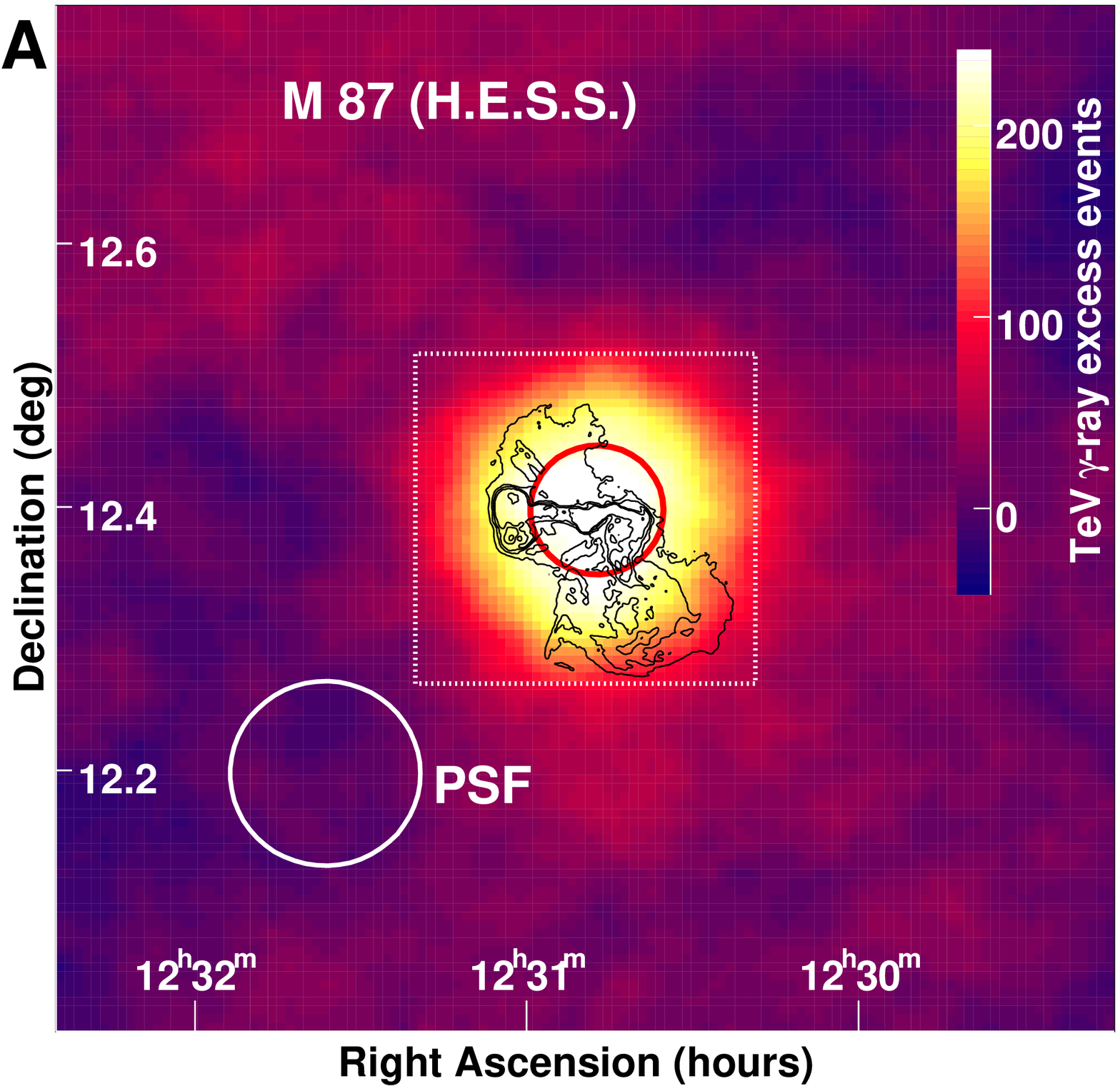,width=0.49\textwidth}
\epsfig{file=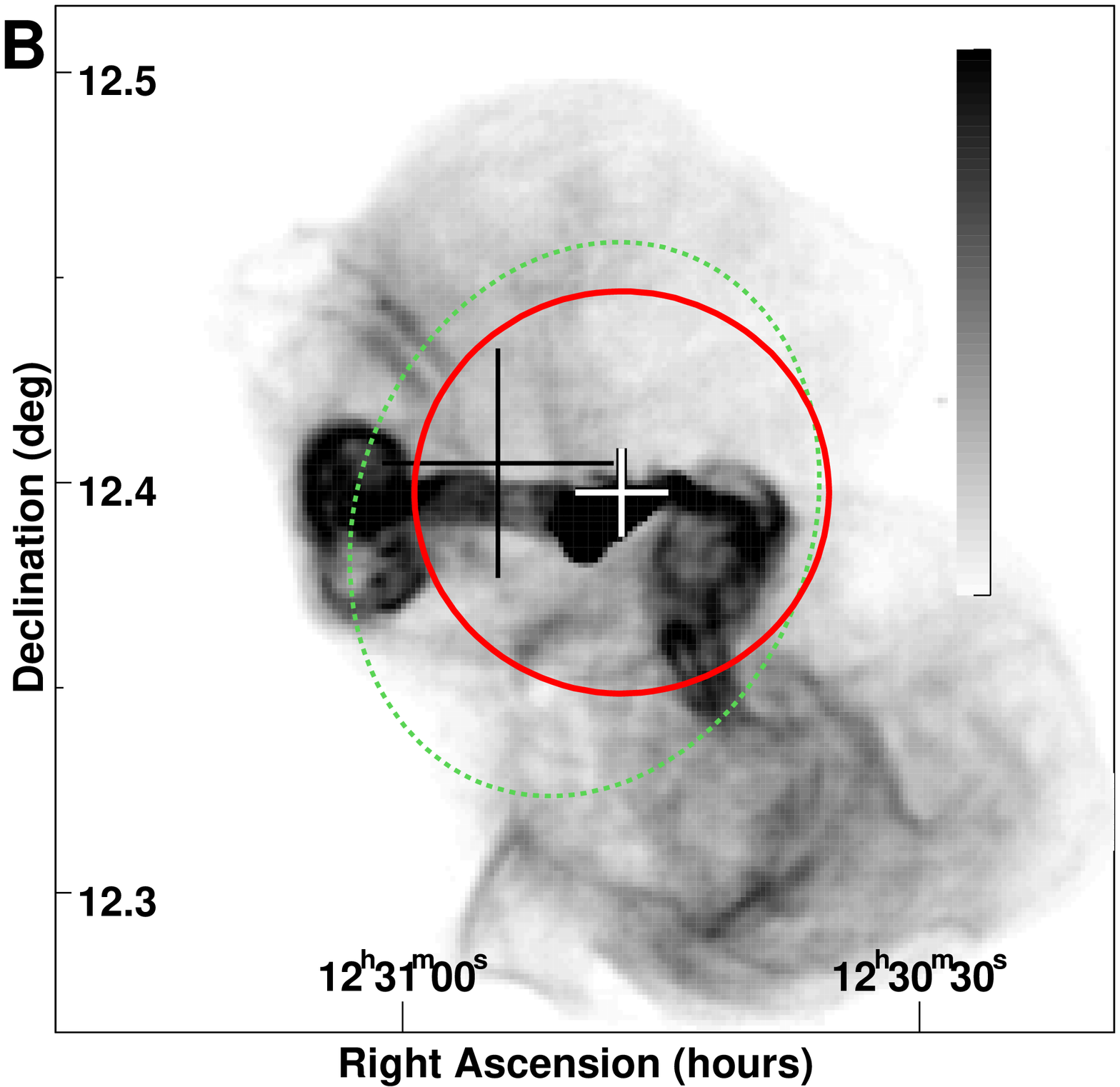,width=0.49\textwidth}

\end{figure}

\noindent {\bf Fig.~1.} Shown are the sky map as well as the position and
extension limit of the TeV $\gamma$-ray emission from M\,87. (A)  
Smoothed TeV $\gamma$-ray excess map (color coded, $0.1^{\circ}$
integration radius) as measured by H.E.S.S. The size ($68\%$ containment
radius) of the H.E.S.S. point spread function (PSF) is also indicated. The
red circle indicates the intrinsic extension upper limit ($99.9\%$
confidence level) of $3$~arcmin of the TeV $\gamma$-ray excess
corresponding to $13.7 \, \textrm{kpc}$ in M\,87. The contour lines show
the $90 \, \textrm{cm}$ radio emission \cite{M87RadioMap}. The white box
marks the cut-out shown in the right panel. (B) The $90 \, \textrm{cm}$
radio data \cite{M87RadioMap} measured with the Very Large Array, together
with the TeV position with statistical and $20''$ pointing uncertainty
errors (white cross) and again the $99.9\%$ c.l. extension upper limit
(red circle). Note that the size of the emission region deduced from the
short-term variability is $\sim 10^{6}$ times smaller. The black cross
marks the position and statistical error of the $\gamma$-ray source
reported by HEGRA. The green ellipse indicates the host galaxy seen in the
optical wavelengths with an extension of $8.3 \times 6.6$ arcmin in
diameter.

\clearpage

\begin{figure}

\epsfig{file=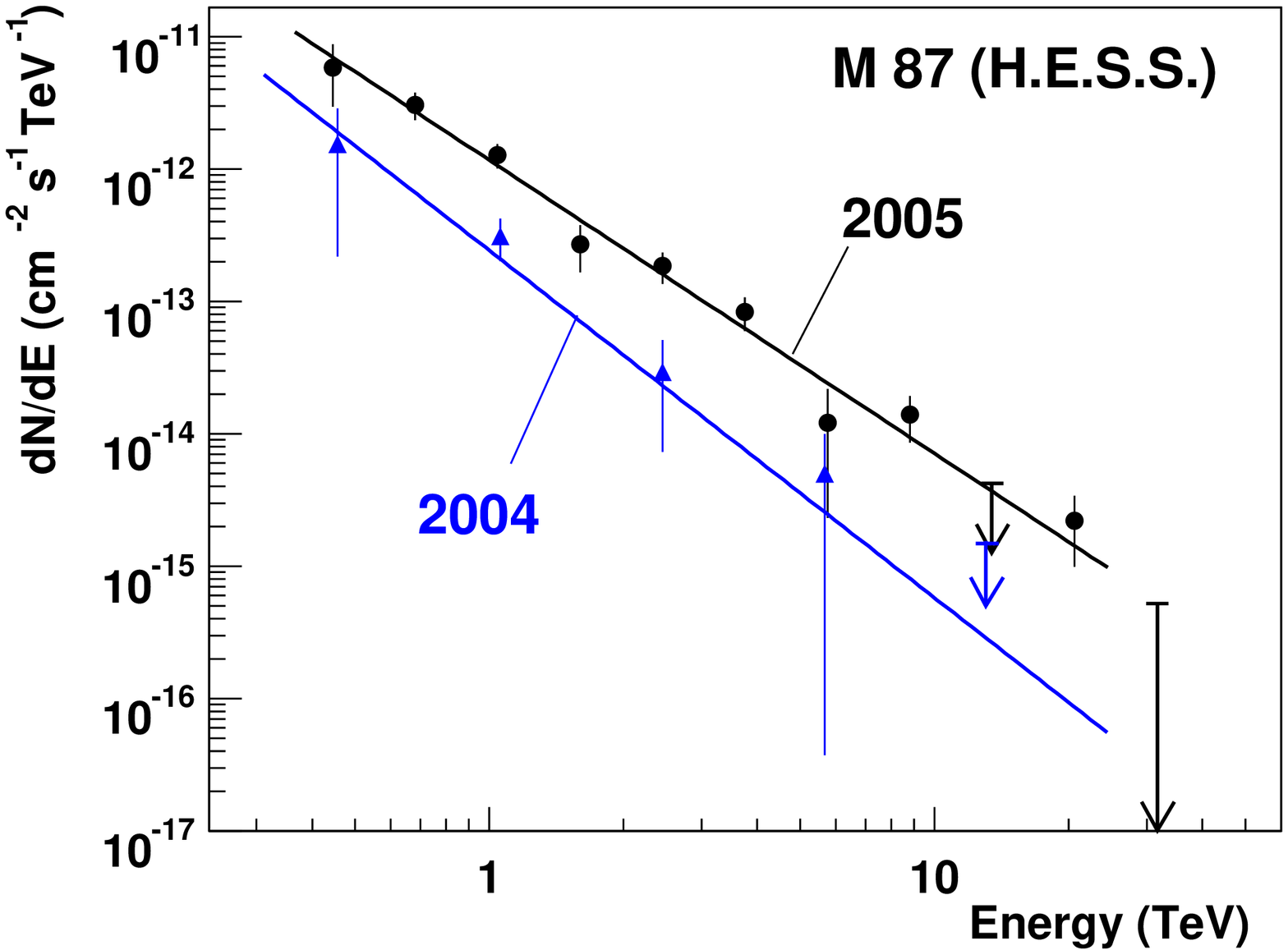,width=0.95\textwidth}

\end{figure}

\noindent {\bf Fig.~2.} The differential energy spectrum of M\,87 obtained
from the 2004 and the 2005 data (using standard event selection cuts
\cite{HESS_HardCuts}) covering a range of $\sim 400 \, \textrm{GeV}$ to
$\sim 10 \, \textrm{TeV}$. Spectra for the 2003 and 2006 data sets could
not be derived due to limited event statistics. Flux points with a
statistical significance less than $1.5$~standard deviations are given as
upper limits ($99.9\%$ c.l.). The corresponding fits of a
power-law function $\textrm{d}N / \textrm{d}E = I_{0} \cdot (E / 1 \,
\textrm{TeV})^{-\Gamma}$ are indicated as lines. The photon indices are
$\Gamma = 2.62 \pm 0.35$ (2004 data) and $\Gamma = 2.22 \pm 0.15$ (2005
data). Aside from the difference in the flux normalisation by a factor of
$\sim 5$ ($I_{0} = (2.43 \pm 0.75) \times 10^{-13} \, \textrm{cm}^{-2} \,
\textrm{s}^{-1} \, \textrm{TeV}^{-1}$ in 2004 and $I_{0} = (11.7 \pm 1.6)
\times 10^{-13} \textrm{cm}^{-2} \, \textrm{s}^{-1} \, \textrm{TeV}^{-1}$
in 2005) no variation in spectral shape is found within errors. The
systematic error on the photon index and flux normalisation are estimated
to be $\Delta \Gamma = 0.1$, and $\Delta I_{0} / I_{0} = 0.2$,
respectively.


\clearpage

\begin{figure}

\epsfig{file=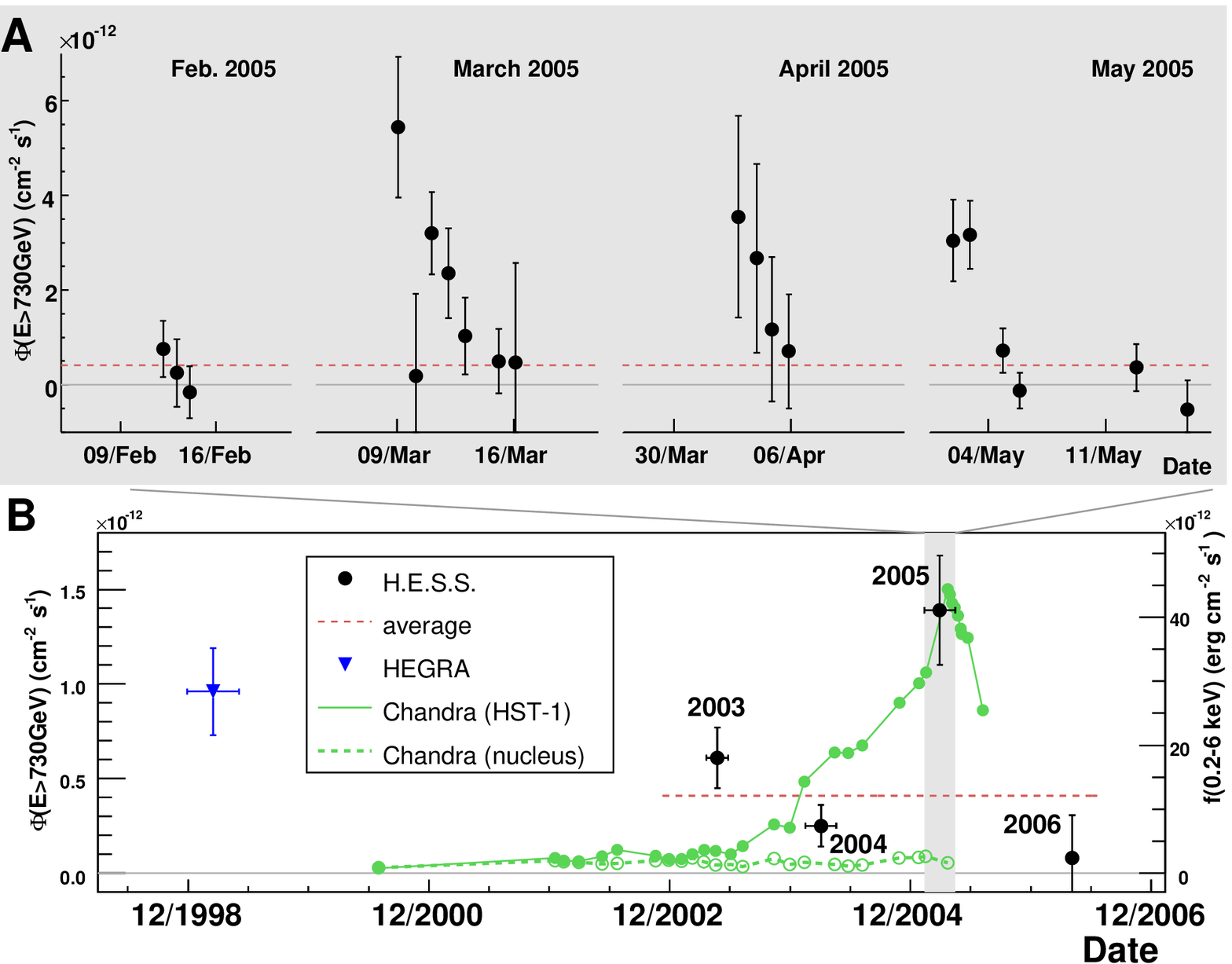,width=0.98\textwidth}

\end{figure}

\noindent {\bf Fig.~3.} Gamma-ray flux above an energy of $730 \,
\textrm{GeV}$ as a function of time. The given error bars correspond to
statistical errors. (B) The average flux values for the years 2003 to 2006
as measured with H.E.S.S. together with a fit of a constant function (red
line). The flux reported by HEGRA is also drawn (a systematic error must
be taken into account when comparing results from the two instruments).  
(A) The night-by-night fluxes for the four individual months (February to
May) of the high-state measurements in 2005 with significant variability
on (flux doubling) time-scales of $\sim 2$~days. The green points in (B)
correspond to the $0.2 - 6 \, \textrm{keV}$ X-ray flux of the knot HST-1
(solid, \cite{HST-1_Xrays}) and the nucleus (dashed,
\cite{ChandraCoreHarris}) as measured by Chandra; the lines are linear
interpolations of the flux points. No unique correlation between the flux
of X-rays and TeV $\gamma$-rays can be identified (the X-ray/TeV data were
not taken simultaneously).

\newpage
\section*{Supporting Materials and methods}

$^\ast$To whom correspondence should be addressed;\\ E-mail:
matthias.beilicke@desy.de, Olivier.Martineau-Huynh@lpnhep.in2p3.fr

\subsubsection*{H.E.S.S. observations and standard analysis}

When a primary $\gamma$-ray photon (or hadron) of TeV energies enters the
earth's atmosphere, an extended air shower consisting of millions of
secondary particles develops in the atmosphere. The Cherenkov light which
is emitted in the air shower is recorded by the photomultiplier camera
situated in the focal plane of each H.E.S.S. Cherenkov telescope. The
recorded images of the shower are parameterized using a set of parameters
(referred to as Hillas parameters) and are used to reconstruct for example
the energy and direction of the primary particle. The geometrical width of
an image is scaled using an expectation value for the corresponding
observation conditions (zenith angle, shower distance, etc.) and is used
to distinguish between $\gamma$-ray and hadron induced showers. This
procedure is referred to as the H.E.S.S. standard analysis and is
described in more detail elsewhere \cite{HESS_HardCuts}. The simultaneous
recording with up to four telescopes (stereoscopic observation) with this
new generation experiment allows an improved (as compared to single
telescope observations)  event-by-event measurement of the direction
($\Delta \theta \sim 0.1^{\circ}$) and energy ($\Delta E / E \approx
15\%$) as well as a superior background suppression of charged cosmic
rays. Hard event selection cuts have been used in the analysis reported
here providing an optimum background rejection for sources with hard
energy spectra at some expense of a lower event rate and an increased
energy threshold of $640 \, \textrm{GeV}$ for the average zenith angle of
$40^{\circ}$ of the M\,87 observations. The energy spectrum was derived
using the standard cuts with a lower threshold of $\sim 400 \,
\textrm{GeV}$ for the same zenith angle.

The H.E.S.S. observations of M\,87 were performed between 2003 and 2006
for a total of $89$ hours after data quality selection cuts (Tab.~S1). The
2003 data ($\sim 25 \, \textrm{h}$) were taken with only two operational
telescopes during the construction phase, while the measurements of the
following years have been performed with the full four telescope array
which has a twice better sensitivity as compared to the two-telescope
setup.  Therefore, the energy spectra and the sky position and extension
limit of the TeV $\gamma$-ray excess were derived from the 2004-2006 data
only. Generating an energy spectrum from the 2006 data was not possible
due to the very limited event statistics.

\subsubsection*{Application of an alternative analysis method}

The results have been cross-checked with a recently developed alternative
analysis method. Beside an independent calibration of the H.E.S.S. raw
data \cite{HESS_Calibration}, this technique is based on the combination
of a parameterisation of the shower images using the moment method of
Hillas (similar to the one described above), and the technique refered to
as model analysis \cite{ModelAnalysis}. This latter method implies a
pixel-by-pixel comparison of the shower images (recorded with the H.E.S.S.
photomultiplier cameras) with a template generated by a semi-analytical
shower development model. The $\gamma$-ray primary energy, direction and
impact position are obtained by maximising a log-likelihood function
associated with this comparison. The model analysis uses all available
pixels in the camera, without the requirement of an image cleaning. Since
the background rejection is based on independent variables in these two
methods, their combination in this alternative analysis improves the
background rejection. The combined analysis used here therefore yields
more significant results than the Hillas parameter based standard analysis
alone.

The light curve of the 2005 flux high state of M\,87, in which the
short-term variability is identified, obtained with this alternative
analysis method is shown in Fig.~S1. The corresponding significance of the
short-term variability is above $6$~standard deviations, clearly
confirming the results obtained from the standard analysis described in
the main text. The energy spectra ($\Gamma = 2.61 \pm 0.24$, $I_{0} =
(3.60 \pm 0.57) \times 10^{-13} \, \textrm{cm}^{-2} \, \textrm{s}^{-1} \,
\textrm{TeV}^{-1}$ for the 2004 data and $\Gamma = 2.20 \pm 0.09$, $I_{0}
= (13.9 \pm 1.2) \times 10^{-13} \, \textrm{cm}^{-2} \, \textrm{s}^{-1} \,
\textrm{TeV}^{-1}$ for the 2005 data) as well as the sky position of
$\alpha = 12^{\textrm{h}} 30^{\textrm{m}} 50.0^{\textrm{s}} \pm
1.3^{\textrm{s}}$ and $\delta = +12^{\circ} 23' 53'' \pm 19''$ (J2000.0)
and extension limit of $3$~arcmin of the TeV $\gamma$-ray excess are also
compatible with the results of the standard analysis.

\subsubsection*{The peculiar knot HST-1}

HST-1 is the innermost resolved knot in the jet of M\,87 (Fig.~S2). Up to
now, no extension or substructure of HST-1 could be resolved in the X-ray
range. This region is the site of many violent events, with X-ray flares
exceeding the luminosity of the core emission and super-luminal blobs
being detected downstream. The radio, UV and X-ray fluxes
\cite{HST-1_Xrays} of HST-1 have increased by more than a factor of $50$
over the period from 2000 to 2005 (Fig.~3). In addition, isolated flares
with variability on time-scales of about $1$~month have been detected both
in optical \cite{M87_JetOptVar} and X-ray \cite{HST-1_Xrays} energy bands.

\clearpage



\epsfig{file=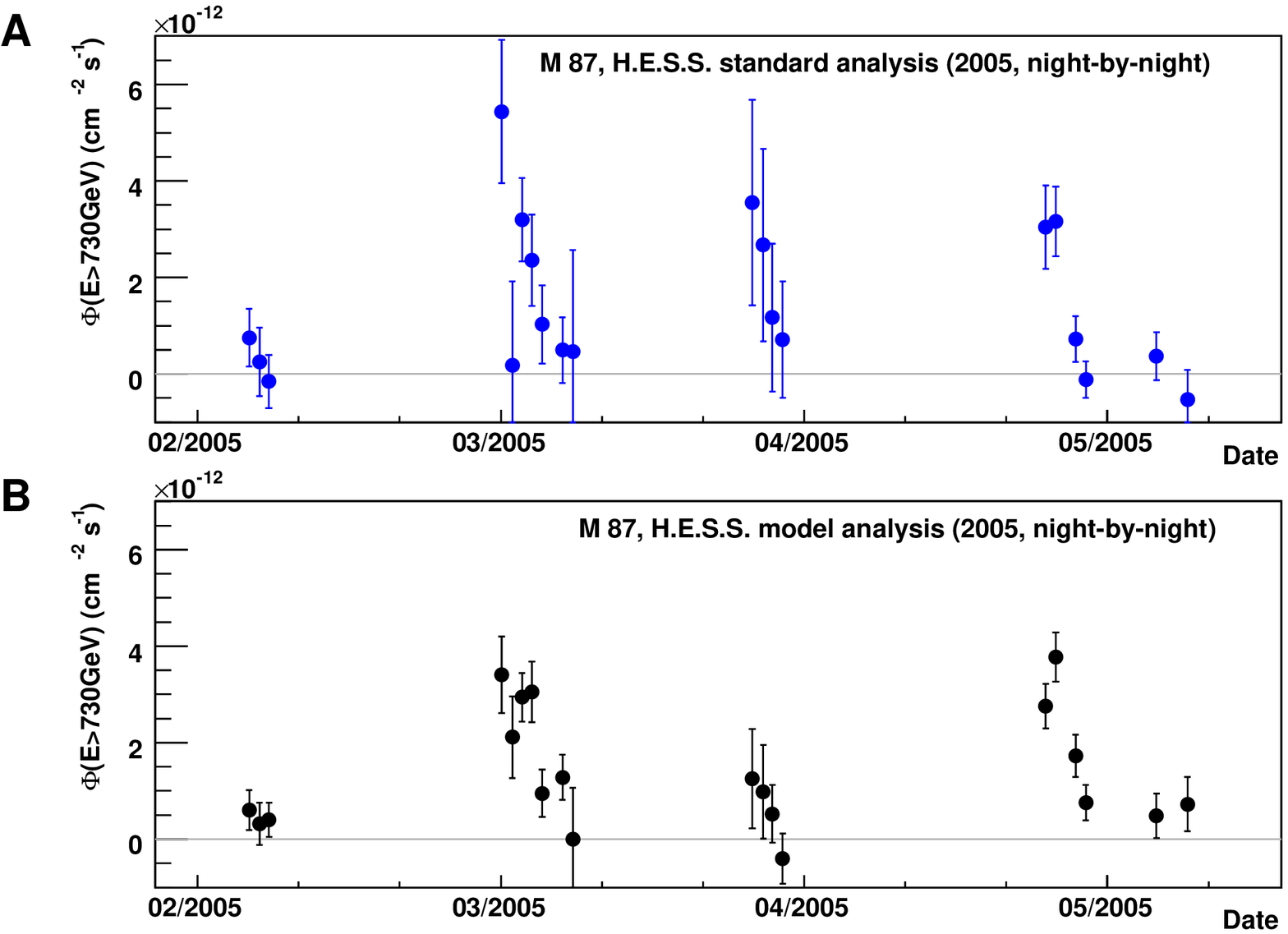,width=0.98\textwidth}


\vspace*{1.0cm}

\noindent {\bf Fig.~S1.} Gamma-ray flux above an energy of $730 \,
\textrm{GeV}$ as a function of time for the 2005 high state (February to
May) from the two analysis methods in comparison. The given error bars
correspond to statistical errors. (A) The results of the standard analysis
shown in Fig.~3, whereas (B) shows the fluxes obtained using the model
analysis \cite{ModelAnalysis}, including an independent calibration chain.
Due to its higher $\gamma$-ray acceptance, about $60\%$ of the
$\gamma$-ray events obtained from the model analysis are not contained in
the corresponding standard analysis in which the hard event selection cuts
were used, making the two data sets (partly) statistically independent.
Both methods exhibit short-time variability with comparable time-scales.


\epsfig{file=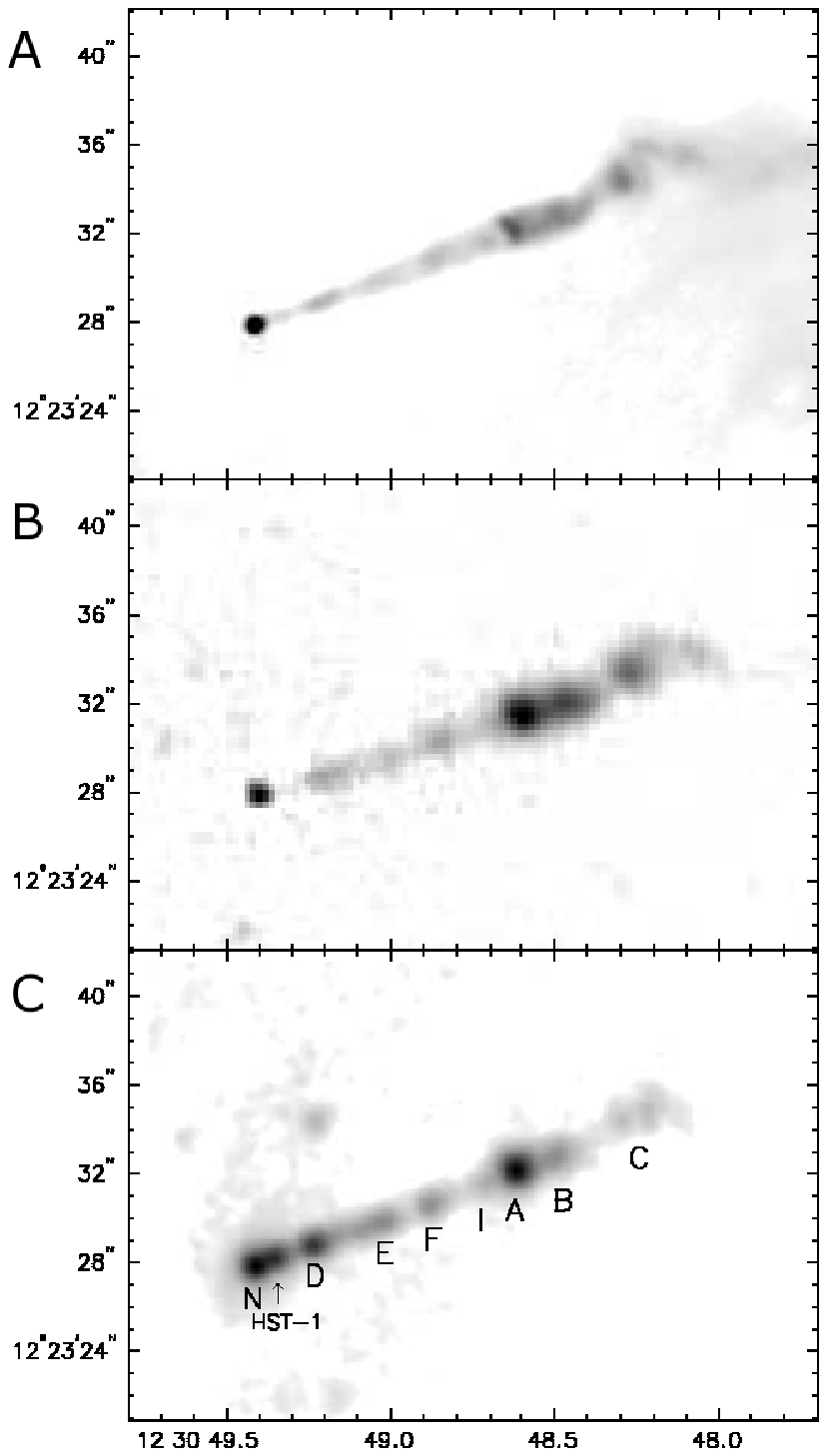,width=0.5\textwidth}

\vspace*{1.0cm}

\noindent {\bf Fig.~S2.} The central $2 \, \textrm{kpc}$ jet of M\,87
which is resolved at radio (A), optical (B) and X-ray (C)  energies. The
nucleus (N), the closest knot HST-1 as well as other knots in the jet are
indicated. The image is adopted from \cite{M87_Jet_Image}.

\clearpage

\noindent {\bf Tab.~S1.} Summary of H.E.S.S. observations of M\,87. Shown
is the time period for each year in which observations were performed, the
observation time $T_{\textrm{obs}}$ and $T_{\textrm{live}}$ (corrected for
the detector dead time) as well as the number of telescopes in the system.
The 2003 data (marked with a $*$) were taken with only two telescopes in a
single operation mode (no stereoscopic trigger) during the construction 
phase.

\vspace*{1.0cm}

\begin{tabular}{llrrr}

Year & time period & $T_{\textrm{obs}}$ [h]
	& $T_{\textrm{live}}$ [h] & $N_{\textrm{tel}}$ \\

\hline

2003$^{*}$ & 24.~Apr. - 28.~June  & $25.1$ & $19.2$ & $2$ \\
2004       & 17.~Feb. - 23.~May   & $35.2$ & $31.4$ & $4$ \\
2005       & 12.~Feb. - 15.~May   & $23.0$ & $21.1$ & $4$ \\
2006       & 25.~Apr. - 18.~May   & $5.3$  & $4.8$  & $4$ \\

\end{tabular}


\bibliography{scibib}

\begin{thebibliography}{3}
\bibitem{SOM} Materials and methods are available as supporting material 
on Science Online.
\bibitem{DefDopplerFactor} Emission from a region which is moving with a
relativistic speed $\beta = v / c$ ($c$ is the speed of light) is boosted
along the direction of movement (relativistic beaming). The boost is a
function of the observation angle $\theta$ relative to this direction and
is described by the Doppler factor $\delta = (\Gamma (1 - \beta \cos
\theta))^{-1}$, where $\Gamma = (1-\beta^{2})^{-1/2}$ is the Lorentz
factor of the emission region.
\end{thebibliography}

\bibliographystyle{Science}

\end{document}